\documentstyle[prd,preprint,tighten,eqsecnum,aps]{revtex}
\begin{document}
\draft
\title{Thermal Two Point Function of a Heavy Muon 
in hot QED plasma within Bloch Nordsieck Approximation}
\author{K.~Takashiba\thanks{Electronic address: 
niegawa@ocux.physics.osaka-cu.ac.jp}}
\address{Department of Physics, Faculty of Science,\\
Osaka City University, Sugimoto, Sumiyoshi-ku, \\
Osaka, 558, Japan}
\date{\today}
\maketitle


\begin{abstract}
 The thermal propagator of a heavy muon propagating in a hot QED 
plasma is examined within the Bloch-Nordsieck approximation, 
which is valid in the infrared region. 
It is shown that the muon damping rate is finite,
in contrast to the lower-order calculation with hard thermal loop 
resummations taken into account.
\end{abstract} 
\pacs{11.10.Wx, 12.20.-m, 12.38.Lg, 13.35.Bv}
\narrowtext

\section{Introduction} 
  Because of much interest in the early Universe 
as well as in the quark-gluon plasma to be produced 
in heavy-ion collisions, many authors have studied 
gauge theories at high temperature. 
Thermal propagators of bosons in real-time thermal field theory 
carry the Boltzmann factor $1/(e^{E/T} -1)$. For a massless
boson like gauge boson, this factor behaves as $\sim T/k$ 
at the infrared region, $k \ll T$, which leads to stronger 
infrared divergences in thermal amplitudes than in vacuum theory.
In vacuum theory, it is well known\cite{Blo,Yen,Mut} that all the physical 
quantities are free from infrared singularities, provided one 
sums over all degenerate final states (Bloch Nordsieck mechanism).
An important exception to this general statement is the so-called
Coulomb singularity.
For thermal reaction rates, due to the stronger infrared 
singularities mentioned above, it is not known if they do not 
possess infrared singularities, except again for Coulomb 
singularities. 
Up until now, a proof is available of absence of leading 
infrared singularities in a generic thermal reaction rate\cite{nie} 
and of all infrared singularities in a bremsstrahlung process within 
some specific approximation\cite{wel}. 

It has been established\cite{Pis0,Pis1,Pis2} that, to obtain a consistent 
perturbation expansion, resummations of hard-thermal-loop (HTL) 
contributions are necessary.  
According to this HTL resummation scheme, infrared behaviors of 
thermal amplitudes are softened or screened compared with those 
in naive perturbative calculation. 
this softening renders some otherwise divergent physical quantities 
finite\cite{Pis1,bur1}.
There are, however, some other physical quantities which are still 
infrared divergent to leading order in HTL resummation scheme.
A well-known example is the quark (muon) damping rate in a 
hot QCD (QED) plasma\cite{bur2}.
As a matter of fact, the quark (muon) damping rate diverges due to 
the Coulomb singularity.
In naive perturbation calculation, the divergence is of power type.  
In HTL resummation scheme, as mentioned above, the infrared behaviour 
is softened and the infrared divergence becomes to be of logarithmic 
type.
It is expected, at least for quark case, that the infrared singularity
is screened when still higher order resummation is performed. 
As far as the infrared behaviour of a spinor field in QED is concerned,
Bloch-Nordsieck approximation is a good approximation\cite{Bog,Ste}. 
In this paper, the thermal two-point function of a heavy muon in a QED
plasma is computed within this approximation. Various properties of
the result are discussed, among those is an especially important 
conclusion: The decaying behavior of a heavy muon does not suffer from
divergence.
 
\section{Thermal Two Point Function of a heavy muon} 

By the term \lq\lq heavy muon\rq\rq we mean so heavy a muon that it is
not thermalized, i.e., $e^{-E/T} \ll 1$. 
The hot QED plasma means that the temperature $T$ of the plasma is 
much higher than the electron mass $m$, $T \gg m$.
As discussed in Sec.~I, the infrared behavior of the theory is 
responsible for the divergent damping rate. 
Thus, we employ the Bloch-Nordsieck approximation to the muon 
sector of the QED Lagrangian.
The Lagrangian density is obtained\cite{Bog,Ste} by substituting 
a constant vector $u^{\mu}$ $(u^2=1)$ for $\gamma^{\mu}$ 
in the muon sector of the standard QED Lagrangian density: 
\begin{eqnarray}              
  {\cal L}^{total} & = & {\cal L}_{\Psi}
                       + {\cal L}_{\psi} + {\cal L}_{A} \;,
         \label{lagtot} \\
  {\cal L}_{\Psi}   & = &
                        \overline{\Psi} (i u \cdot \partial 
                                + e u \cdot A -M ) \Psi \;,
         \label{lag-muon} \\
  {\cal L}_{\psi}   & = &
                        \overline{\psi} (i \gamma \cdot{\partial} 
                                + e \gamma \cdot A -m ) \psi \;,
         \label{lag-elec} \\
  {\cal L}_{A}      & = &
                        -\frac{1}{4} F_{\mu \nu} F^{\mu \nu} \;,
         \label{lag-f} \\
  F_{\mu \nu}      & = & 
                        \partial_{\mu} A_{\nu}
                                - \partial_{\nu} A_{\mu} \;,
         \label{f-strength} 
\end{eqnarray}
where $\Psi, \psi, M, m$, and $A$ stand, respectively, for 
heavy muon field, electron field , muon mass, electron mass 
and photon field. 
It is known that, in the infrared region, the spinor structure 
of the heavy muon does not play any significant roles\cite{Bog,Ste}. 
Then the above Lagrangian density well approximates the exact QED 
Lagrangian in the infrared region.    

To obtain the thermal two-point function of the heavy muon, 
we start from the Euclidean or imaginary-time formalism of thermal 
field theory, initiated by Matsubara\cite{Mat}, in which a time arrow 
flows from zero to $-i/T$ in a complex time plane.
In this formalism, the energy is discrete; 
$p_{4} = 2 \pi n T$ $(n = 0, \pm 1, \pm 2, \cdots)$ for a boson field 
and $p_{4} = \pi (2n+1) T$ for a fermion field. 
At the final stage, we analytically continue the two-point function, 
thus obtained, to real time (or real energy). 
A four vector $a^{\mu}=(\vec{a},a_{4})$ in the imaginary-time 
formalism is obtained from the counter part $(a_{0}, \vec{a})$ in 
Minkowski space by an analytic continuation, 
$a_{0} \longrightarrow -i a_{4}$.
Similar continuation is made for tensors.

The generating functional is introduced as usual,
\begin{eqnarray}                                
        Z & = & 
                \int {\cal D} \overline{\Psi}
                     {\cal D} \Psi
                     {\cal D} \overline{\psi}
                     {\cal D} \psi 
                     {\cal D} A
                            \exp{S} \;,
       \label{totgene} \\
        S 
          & = & 
                \int_{0}^{\beta} d^{\,4}x 
                 \biggl(
                  {\cal L}^{total} 
                  - \frac{1}{2 \lambda}
                  \left( \partial \cdot A \right)^2
        \nonumber \\
          &   &
                  + \overline{\eta}  \Psi + \overline{\Psi} \eta
                  + \overline{\zeta} \psi + \overline{\psi} \zeta
                  + j \cdot A
                 \biggr)
                  \;\;\; ,
       \label{totact} 
\end{eqnarray}
where the covariant gauge is employed.
In Eq. (\ref{totact}), $\eta, \zeta, j, \cdots$ are sources 
being conjugate to the fields with which they couple.
After integrating over $\overline{\Psi},\Psi,\overline{\psi}$, and $\psi$ 
in Eq. (\ref{totgene}), 
the two point function of the heavy muon is obtained as, 
\begin{eqnarray}                               
        G_{E}  & = & 
                     \left. 
                      - \frac{1}{Z}
                        \frac{\delta}{\delta \overline{\eta}(x)}
                        \frac{\delta}{\delta \eta(y)}
                        \; Z[\eta,  \overline{\eta},
                              \zeta, \overline{\zeta},
                              j ] \right|_{\eta,\cdots = 0}
       \nonumber \\      
               & = & 
                        \frac{1}{Z}
                        \int {\cal D}A
                        \det(i u_{E}  \cdot \partial -e u_{E} \cdot A - M)
       \nonumber \\      
               &   & 
                        \times
                        \det(i \gamma \cdot \partial 
                                                -e \gamma \cdot A - m)
                     G(x-y;eA,M) 
       \nonumber \\     
               &   &  
                     \times 
                     \exp \left[ 
                              -\frac{1}{4} \int d^{\,4} x 
                               F_{\mu \nu} F_{\mu \nu} 
                               - \frac{1}{2 \lambda}
                               \left( \partial \cdot A \right)^2
                          \right] \;,
       \label{green}
\end{eqnarray}
where
\begin{eqnarray}
        (i u_{E} \cdot \partial - e u_{E} \cdot A - M)
        G(x-y;eA,M)
               & \equiv &
                   \delta_{E}(x-y) \,,
       \nonumber \\     
       \label{equ} 
\end{eqnarray}
\begin{eqnarray}
        \delta_{E}(x-y) 
               & \equiv & 
                     T \sum_{k_{4} = odd} 
                     \left. 
                          \int \frac{d^3k}{(2\pi)^3}
                          \exp[i k \cdot (x-y)] 
                     \right.
       \label{ope6} \;\;\;, \\
        k_{4}  & = & \pi (2n+1) T
                     \mbox{\hspace*{7ex}} 
                     (n = 0, \pm1, \pm2, \cdots)
                     \;\;\;.
       \label{ope8} 
\end{eqnarray}
In Eq. (\ref{ope6}), \lq\lq$k_{4} = odd$\rq\rq $\;$indicates that the 
summation runs over $n$ as defined in Eq. (\ref{ope8}).

Since the heavy muon is not thermalized, the muon determinant 
has been reduced to $\det (iu_{E} \cdot \partial - M)$ as in vacuum 
theory \cite{Bog}.
Using the formula $\det a = \exp({\mathrm Tr} \ln a)$ and Furry's 
theorem\cite{Itz}, we have
\widetext
\begin{eqnarray}                              
        \det(i \gamma \cdot \partial - e \gamma \cdot A- m) 
               & = & 
                     \det(i \gamma \cdot \partial - m) 
                     \exp \biggl[ 
                          - \frac{e^2}{2} {\mathrm Tr} 
                          \int d^{\,4} {z_{1}} d^{\,4} {z_{2}}
                          (\gamma \cdot A)
                          G_{0}(z_{1} -z_{2};m)
      \nonumber \\
               &   & 
                     \times 
                          (\gamma \cdot A)
                          G_{0}(z_{2} -z_{1};m)
                          + O(e^4) 
                          \biggr] \;,
       \label{det} \\
        G_{0}(x-y;m)
               & = & 
                     \left. 
                      - T \sum_{k_{4} = even}
                      \int \frac{d^{\, 3} k}{(2 \pi)^3}
                      \frac{ 
                           \exp[ik\cdot (x-y)]
                           }{ \gamma \cdot k + m}
                     \right.  
       \label{prop} \;.
\end{eqnarray}
\narrowtext
Here $G_{0}(x-y;m)$ denotes a bare electron propagator.
As seen from Eq. (\ref{green}) with Eq. (\ref{det}),
in the heavy-muon two-point function, the contributions of electrons 
enter through thermal electron loops.
Here we employ the HTL resummation scheme for the electron sector: 
To leading order, among such electron loops, only HTLs that yield 
corrections to soft-photon external lines are important.
In hot QED considered here, the $O(e^4)$ term in Eq. (\ref{det}) 
vanishes\cite{Pis2} to leading order in the HTL resummation scheme.
We then ignore it in the following.

Following \cite{Bog}, we introduce the integral representation,
\begin{eqnarray}
        G(x-y;eA,M)
               & = &
                     -i \int_{0}^{\infty} d\tau
                     \exp \biggl[ 
                                 i \tau ( u_{E} \cdot \partial
      \nonumber \\
               &   & 
                                 - e u_{E} \cdot A - M 
                                 + i \epsilon_{c} ) 
                          \biggr]
                     \delta_{E}(x-y)
       \label{ope} \;,
\end{eqnarray}
where $\epsilon_{c}$ is a ``convergence\rq\rq $\;$factor.
Defining the function $U(\tau)$ as, 
\begin{equation}
        U(\tau)  \equiv 
                     \exp \biggl[ 
                                 i\tau (i u_{E} \cdot \partial
                                        - e u_{E} \cdot A
                                        - M + i \epsilon_{c}) 
                          \biggr]
                     \delta_{E}(x-y)
       \label{ope3} \;,
\end{equation}
\begin{equation}
        G(x-y;eA,M)
                 \equiv
                     - i \int_{0}^{\infty} d\tau \; U(\tau)
       \label{ope2} \;,
\end{equation}
we have
\begin{eqnarray}
        -i \frac{\partial U}{\partial \tau} 
               & = &
                     (i u_{E} \cdot \partial
                     - e u_{E} \cdot A - M 
                     + i \epsilon_{c}) U(\tau)
       \label{ope4} \;\;\;, \\
        U(0)   & = & \delta_{E}(x-y)
       \label{ope5} \;\;\;.
\end{eqnarray}
Now we assume the following form for the solution to 
Eq. (\ref{ope4}) with Eq. (\ref{ope5}), 
\begin{eqnarray}
        U(\tau)  
               & = & 
                    T \sum_{q_{4} = odd} 
                    \int \frac{d^3 q}{(2 \pi)^3}
                    \exp \biggl[ 
                              i K(x-y|A, \tau) 
         \nonumber \\
               &   &         
                              + i q \cdot (x-y) 
                              - i \tau (M + u_{E} \cdot q 
                              -i\epsilon_{c}) 
                         \biggr]
       \label{ope11} \;\;\;.
\end{eqnarray}
Eq. (\ref{ope4}) with Eq. (\ref{ope11}) leads to
\begin{eqnarray}
        -\frac{\partial K}{\partial \tau} 
               & = &
                     u_{E} \cdot \frac{\partial K}{\partial x} 
                     + e u_{E} \cdot A
         \label{ope9} \;, \\
        K(\tau = 0) 
               & = & 0 
         \label{ope10} \;,
\end{eqnarray}
where $u_{E}=(\vec{u},u_{4})$.
It is straightforward to solve this equation in the form: 
\widetext
\begin{eqnarray}
        K(x-y|A, \tau)  
               & = &
                     - e T \sum_{k_{4} = even} 
                     \int \frac{d^3 k}{(2 \pi)^3} 
                     \exp[i k \cdot (x-y)] A(k) \cdot u_{E} 
                     \int_{0}^{\tau} d\tau' 
                     \exp[- i u_{E} \cdot k \tau'] 
       \label{sol}  \;,
                  \\
      k_{4}    
               & = &
                    2 \pi n T 
                    \;\;\;\; (n = 0, \pm1, \pm2, \cdots)  
       \label{sol2} \;,  
\end{eqnarray}
where \lq\lq$k_{4} = even$\rq\rq$\;$indicates that the summation runs over 
$n$ as defined in Eq. (\ref{sol2}).  

Using all the formulae displayed above, and carrying out the integration
by part, we obtain for $G_{E}$ in Eq. (\ref{green}):
\begin{eqnarray}
        G_{E}  & = & 
                    \left.  
                        i \int {\cal D}A \int_{0}^{\infty} d \tau 
                        T \sum_{q_{4} = odd}
                        \int \frac{d^3 q}{(2 \pi)^3}
                    \right.
                        \exp\Biggl[ 
                                 i q \cdot (x-y) 
                                 - i\tau (M +u_{E}\cdot q 
                                 -i\epsilon_{c})
      \nonumber \\
               &   &  
                                 -\frac{e^2}{2} 
                                 \int d^{\, 4} z_{1} d^{\, 4} z_{2} 
                                 {\mathrm Tr}\;
                                 (\gamma \cdot A)
                                 G(z_{1} - z_{2};e=0,m)
                                 (\gamma \cdot A)
                                 G(z_{2} - z_{1};e=0,m)
      \nonumber \\
               &   &
                                 + \int d^{\, 4} z
                                 \frac{1}{2} A^{\mu} 
                                 \left\{
                                     {\delta}^{\mu \nu} 
                                     \partial^2
                                     - \left( 1- \frac{1}{\lambda} \right)
                                     \partial^{\mu} \cdot \partial^{\nu} 
                                 \right\} {A}^{\nu}
      \nonumber \\
               &   &  
                                 - i e T \sum_{k_{4} = even}
                                 \int \frac{d^{\,3} k}{(2 \pi)^3}
                                 A(k) \cdot u_{E} 
                                 \exp[i k \cdot (x-y)]
                                 \int_{0}^{\tau} d\tau\rq 
                                 \exp[- i u_{E} \cdot k \tau\rq]
                            \Biggr] \;.
      \nonumber \\
       \label{G3} 
\end{eqnarray}
Going to the momentum space,
\begin{equation}
        {\cal G}_{E}  
                =  
                    \int d^{\, 4} (x-y) \exp[-i p \cdot (x-y)] G_{E}
       \label{G4} \;,
\end{equation}
we have,
\begin{eqnarray}
        {\cal G}_{E} 
               & = &
                    \left. 
                        i \int {\cal D} A \int_{0}^{\infty} d \tau
                        \exp \Biggl[ 
                        - i \tau (M - u_{E} \cdot p -i \epsilon_{c})
                    \right. 
      \nonumber \\
               &   &  
                    \left.
                        - \frac{T}{2} 
                        \sum_{l_{4} = even} \int \frac{d^{\,3} l}{(2\pi)^3}
                        \;
                        {A}^{\mu}(l) 
                        \left\{ 
                             {\delta}^{\mu \nu} l^2
                             - \left( 1 -\frac{1}{\lambda} \right) 
                             l^{\mu} l^{\nu} 
                             + \Pi^{\mu \nu}(l) 
                        \right\} 
                        {A}^{\nu}(-l) 
                   \right. 
      \nonumber \\
               &   &  
                    \left.
                        - i e T
                        \sum_{l_{4}= even } \int \frac{d^{\,3} l}{(2\pi)^3}
                        \exp[ i l \cdot u_{E} \tau] A(l) \cdot u_{E}
                        \int_{0}^{\tau} d \tau' 
                        \exp[ -i l \cdot u_{E} \tau\rq ]
                    \right. 
                            \Biggr]
       \label{G5} \;,
\end{eqnarray}
where $\Pi^{\mu \nu}$ is the thermal vacuum polarization tensor of the photon:
\begin{eqnarray}
        \Pi^{\mu \nu}(l) 
               & = &
                    \left.
                        e^2 T \sum_{k_{4} = odd} 
                        \int \frac{d^3 k}{(2 \pi)^3}
                        {\mathrm Tr} 
                        {\gamma}^{\mu} 
                        \frac{1}{ \gamma \cdot (k-l) + M }
                        {\gamma}^{\nu} 
                        \frac{1}{ \gamma \cdot k + M }
                    \right.
      \nonumber \\
               & = &
                    \left(
                        \hat{\delta}^{\mu \nu} 
                        - \frac{\kappa^{\mu} \kappa^{\nu}
                               }{
                                 \kappa^2
                                }     
                    \right)
                    \Pi_{T}
                    + 
                    \left(
                        \delta^{\mu 0} \delta^{\nu 0}
                        + \frac{\kappa^{\mu} \kappa^{\nu}
                               }{
                                 \kappa^2
                                }                                 
                        - \frac{l^{\mu} l^{\nu} }{l^2}
                    \right) 
                    \Pi_{L} 
       \label{pol} \;,
\end{eqnarray}
where $\hat{\delta}^{\mu \nu} = \text{diag} \,(1,1,1,0)$ 
and $\kappa^{\mu} = (\vec{l},0 \,)$.
In Eq. (\ref{pol}), $\Pi_{T} \;(\Pi_{L})$ is the transverse (longitudinal) 
component of $\Pi^{\mu \nu}$ (See Appendix A).

The exponent in Eq. (\ref{G5}) is quadratic with respect to 
$A$, and we can perform the functional integration over $A$:
\begin{eqnarray}                                             
        {\cal G}_{E}  
               & = & 
                    i \int_{0}^{\infty} d\tau 
                    \exp\Biggl[ 
                            - i\tau (M + u_{E} \cdot p -i \epsilon_{c})
                            + \frac{\alpha}{2\pi^{2}} 
                            D(\tau, u_{E})
                        \Biggr]
      \label{G6} \;, \\
        D(\tau, u_{E}) 
               & = &  
                    - (2\pi \mu)^{3-D} T \sum_{l_{4} = even}
                    \int d^{\, D}l \; 
                    \frac{1}{ ( l \cdot u_{E} )^2 } 
                    \biggl\{ 1- \cos( l \cdot u_{E} \tau) \biggr\}
      \nonumber \\ 
               &   &
                    \times 
                    \Biggl[ 
                        \frac{1}{ l^2 + \Pi_{T} }
                        \left\{ 
                            {\vec{u}}^2 
                            - \frac{(\vec{l} \cdot \vec{u})^2
                                   }{{\vec{l}}^2} 
                        \right\}
                        + \frac{1}{l^2 + \Pi_{L}}
                        \left\{ 
                            {u_{E}}^2  - \frac{(l \cdot u_{E})^2}{l^2}
                            - {\vec{u}}^2 
                            + \frac{ {(\vec{l} \cdot
                                     \vec{u})}^2}{{\vec{l}}^2}
                        \right\} 
      \nonumber \\ 
               &   &
                        + \lambda 
                        {\left( 
                             \frac{l \cdot u_{E}}{l^2} 
                        \right)}^2 
                    \Biggr] \;,
       \label{G7} 
\end{eqnarray}
\narrowtext
where $\alpha$ is the fine structure constant.
Here, in order to regulate the ultraviolet divergence, we have 
employed the dimensional regularization, $D = 3 - \epsilon$ 
with $ \epsilon > 0 $. (A dimensionful parameter $\mu$ is introduced, 
as usual, so that $e$ remains dimensionless.)
In the HTL approximation, which we employ for the electron sector, 
the explicit form of $\Pi_{T}$ and $\Pi_{L}$ are 
known \cite{Pis0,Pis1,bur1,bur2,wel1}:
\begin{eqnarray}                                            
        \Pi_{T} 
              & = & 
                   - \frac{1}{2} {m_{sc}}^2 \frac{{l_{4}}^2}{{\vec{l}}^2}
                     \left\{
                           1 - \left( 
                                   1+ \frac{{\vec{l}}^2}{{l_{4}}^2} 
                               \right)              
                           \frac{i l_{4}}{2|\vec{l}|}
                           L(i l_{4},|\vec{l}|) 
                     \right\}
       \label{pi-t}  \;, \\
        \Pi_{L} 
              & = &  
                   {m_{sc}}^2 \frac{l^2}{{\vec{l}}^2}
                     \left\{
                           1- \frac{i l_{4}}{2|\vec{l}|} 
                           L(i l_{4},|\vec{l}|)
                     \right\} 
       \label{pi-l}  \;,
\end{eqnarray}
where
\begin{eqnarray}
        m_{sc} 
              & = & 
                   \frac{eT}{\sqrt{3}} 
       \label{mass} \;, \\ 
        L(i l_{4},|\vec{l}|) 
              & = & 
                   \ln \frac{i l_{4} + |\vec{l}|}{i l_{4} - |\vec{l}|}
       \label{legendre} \;.
\end{eqnarray}
$m_{sc}$ in Eq. (\ref{mass}) is thermal or 
Debye mass of the thermal photon.
In obtaining Eqs. (\ref{G7}) -- (\ref{legendre}), 
the electron mass $m$ has been ignored (cf. the first paragraph). 

Let us turn back to $D(\tau, u_{E})$ in Eq. (\ref{G7}).
After a straightforward but tedious calculation (cf. Appendix A), 
we have 
\widetext
\begin{eqnarray}
        D(\tau, u_{E})
              & = & 
                   - 2  \pi^2 T \tau \frac{{u_{4}}^2}{|\vec{u}|} 
                   \left(
                       \frac{1}{\epsilon}
                       + \ln \frac{2\pi \mu }{\sqrt{\pi} m_{sc}} 
                       - \frac{\gamma}{2}    
                       - \frac{1 - e^{-m_{sc} |\vec{u}| \tau}
                              }{
                             |\vec{u}| \tau m_{sc}
                               }
                       + E_{i}(- |\vec{u}| \tau m_{sc})
                   \right)   
     \nonumber \\              
              &   & 
                   - 2 \pi^2 T \tau |\vec{u}| 
                   \left( 
                         \frac{1}{\epsilon} 
                         + \frac{\gamma}{2} -1
                         + \ln \frac{2 \pi \mu |\vec{u}| \tau
                                    }{\sqrt{\pi}}
                   \right)  
                   - (\lambda -1) \pi^2 T \tau  |\vec{u}|
     \nonumber \\              
              &   &   
                    + 2\pi^2 T \tau\frac{{u_{E}}^2}{|\vec{u}|} 
                    \left(
                       \frac{1}{\epsilon} -\frac{\gamma}{2}
                       -\ln \sqrt{\pi} 
                    \right)
                    + \frac{2\pi}{\epsilon} +\pi \gamma
                    - 2\pi \ln \frac{4\pi^{3/2} T}{2\pi \mu}
                    - \frac{2\pi u_{4}}{|\vec{u}|} 
                    \tan^{-1} \frac{u_{4}}{|\vec{u}|} 
      \nonumber \\              
              &   & 
                       - 2 \pi^2 T \tau \frac{{u_{E}}^2}{|\vec{u}|}
                       \left(
                           \ln \frac{T}{2\pi \mu}
                           + \ln \frac{u_{E}}{|\vec{u}|} 
                       \right)
                       + 4 \pi^2 T \tau \frac{{u_{E}}^2}{|\vec{u}|}
                       \int_{1}^{\infty} d x \frac{1}{x}
                       \frac{1}{
                              e^{2\pi T |\vec{u}| \tau x} -1 
                               }
      \nonumber \\              
              &   & 
                       + \frac{ \pi {u_{E}}^2 
                              }{  |\vec{u}| 
                             (|\vec{u}|  -i u_{4}) 
                              } 
                       \ln (1 - 
                          e^{-2 \pi T \tau (|\vec{u}| -i u_{4})} 
                           ) + c.c.
      \nonumber \\              
              &   & 
                       + 2 \pi^2 T \tau 
                       \frac{u_{4} {u_{E}}^2}{|\vec{u}|} 
                       \left(
                           i \int_{0}^{1} d x
                           \frac{1}{
                             |\vec{u}| - i u_{4} x  
                                   }
                           \frac{1}{
                              e^{2 \pi T \tau (|\vec{u}| 
                                  - i u_{4} x )
                                }
                               - 1 }
                           + c.c.
                       \right)
      \nonumber \\              
              &   &
                   + \pi (\lambda -1) 
                   \left[
                       - \frac{1}{ \epsilon} 
                       - \ln \frac{2 \pi\mu}{\sqrt{\pi}}
                       + \gamma
                       + (- \gamma + \ln 4 \pi T)
                       - 2 \ln (1 -e^{-2 \pi T |\vec{u}| \tau })
                  \right.
      \nonumber \\              
              &   &
                       - 2 |\vec{u}| \tau \pi T
                       \frac{1}{e^{ 2 \pi T |\vec{u}| \tau} -1}
                       + \frac{1}{2} \ln ( 1 
                          - e^{- 2 \pi T \tau 
                          (|\vec{u}| - i u_{4}) 
                              }           
                                         ) 
                       + c.c.
      \nonumber \\              
              &   &
                       + \pi T \tau
                          (|\vec{u}| - i u_{4}) 
                          \frac{1}{
                            e^{2 \pi T \tau 
                              ( |\vec{u}| - i u_{4} ) -1 }
                                  }
                       + c.c.
      \nonumber \\              
              &   &
                       + 3 i u_{4} \pi T \tau
                       \int_{0}^{1} dx
                       \frac{1}{
                          e^{2 \pi T \tau 
                          (|\vec{u}| -i u_{4} ) 
                            } -1 
                                }
                       + c.c.
      \nonumber \\              
              &   &
                       - i 2 \tau^2 \pi^2 T^2 u_{4}                            
                       \int_{0}^{1} dx
                       (|\vec{u}| -i u_{4} x) 
                       \frac{
                        e^{ -2 \pi T \tau
                        (|\vec{u}| -i u_{4} x)  } 
                           }{
                        ( e^{ -2 \pi T \tau
                        (|\vec{u}| -i u_{4} x) } -1 
                        )^2
                            }
                       + c.c.
                   \Biggr] \;, 
       \label{detail}
\end{eqnarray}
where $\gamma \sim 0.57 \cdots$ and $E_{i}(-|\vec{u}| m_{sc} \tau)$ 
are, respectively, the Euler constant and 
the exponential integral function and 
$u_{E}= \sqrt{{u_{4}}^2 + {|\vec{u}|}^2 }$.
The details are given in Appendix A. 

An analytic continuation of ${\cal G}_{E}$ in Eq. (\ref{G6}) 
with Eq. (\ref{detail}) back to Minkowski space, which is 
obtained through $p_{4} \longrightarrow i(p_{0} + i \epsilon), \;
u_{4} \longrightarrow i (u_{0} + i \epsilon)$, yields the retarded 
function. 
We normalize it through MS scheme\cite{Mut}, which amounts to choose 
the renormalization factor $Z_{2}$ as
\begin{equation}
        Z_{2} = 
        e^{\alpha 
        \frac{3 - \lambda }{2 \pi} 
        \frac{1}{\epsilon}} 
       \label{sub} \;.
\end{equation}
Then we obtain,
\begin{equation}
        {\cal G}_{R}  
                =   i \int_{0}^{\infty} d \tau
                    \exp{ \left\{ 
                                 - i \tau ( M - u \cdot p -i \epsilon_{c} ) 
                                 + \frac{\alpha}{2\pi^2} {\cal D}_{R} 
                          \right\} 
                        } \;\;,
       \label{g-r} 
\end{equation}
where $u \cdot p = u^{\mu} p_{\mu}$ is the scalar product in Minkowski
space.


\section{Physical contents of the two point function}

We study the behaviors of ${\cal D}_{R}$ in Eq. (\ref{g-r}) 
in various limits of parameters involved.
We divide the region $0 \leq \tau < \infty$ into several regions,
which are discriminated by the two parameters 
$\tau_{A}=\frac{1}{|\vec{u}|T}$, 
$\tau_{B}=\frac{1}{m_{sc}|\vec{u}|}$ (see Appendix A).

\noindent a)  $\tau \ll \tau_{A}$
\begin{equation}
        {\cal D}_{R}
                \simeq  
                    2\pi \left\{ 
                             1 
                             + \frac{3 - \lambda}{2} 
                             \left(
                                 \frac{\gamma}{2} 
                                 + \ln \frac{ \mu \tau }{\sqrt{\pi}}
                                 - \frac{i \pi}{2} 
                             \right)
                         \right\}
       \label{R1} 
\end{equation}
b)  $ \tau_{A} \ll \tau  \ll \tau_{B} $
\begin{eqnarray}
        {\cal D}_{R}  
              & \simeq &
                   2\pi^2 T \tau \left\{
                                     \frac{1}{|\vec{u}|} 
                                     \left( \ln{\tau T } 
                                        - \frac{i\pi}{2} + \gamma -1 
                                     \right) 
                                     - \frac{|\vec{u}|}{2} ( \lambda -1)  
                                  \right\}
        \nonumber \\
              &        &
                   + 2 \pi \frac{3 - \lambda}{2}
                   \left(
                       \frac{\gamma}{2} 
                       - \ln \frac{2 \sqrt{\pi} T}{\mu}
                   \right)
                   - \pi\frac{u_{0}}{|\vec{u}|} 
                    {\ln \frac{u_{0}-|\vec{u}|}{u_{0}+|\vec{u}|}}   
                   -i\pi^2 \frac{u_{0}}{|\vec{u}|}
       \label{R2} 
\end{eqnarray}
c)  $ \tau_{B} \ll \tau $
\begin{eqnarray}
        {\cal D}_{R} 
              & \simeq &
                   -2 \pi^2 T \tau |\vec{u}| 
                     \left\{ \ln{\tau T } 
                             - \frac{i\pi}{2} + \gamma -1
                             - \frac{{u_{0}}^2}{|\vec{u}|^2}
                             \left( \ln{ \frac{T}{|\vec{u}|m_{sc}}} 
                                    - \frac{i\pi}{2} 
                             \right)
                             + \frac{\lambda -1}{2} 
                     \right\} 
        \nonumber \\
              &        &
                   -2 \pi^2 T 
                   \frac{{u_{0}}^2}{|\vec{u}|^2} \frac{1}{m_{sc}}
                   + 2 \pi \frac{3 - \lambda}{2}
                   \left(
                       \frac{\gamma}{2} 
                       - \ln \frac{2 \sqrt{\pi} T}{\mu}
                   \right)
                   - \pi\frac{u_{0}}{|\vec{u}|} 
                    {\ln \frac{u_{0}-|\vec{u}|}{u_{0}+|\vec{u}|}}   
                   -i\pi^2 \frac{u_{0}}{|\vec{u}|} \;\;.
       \label{R3}
\end{eqnarray}
In the above equations, use has been made of $u^2=1$. 

When the muon is far from the mass shell 
$|M - p \cdot u| \gg M$, in Eq. (\ref{g-r}), 
the region $\tau  \ll \tau_{A}$ gives the dominant contribution:  
\begin{eqnarray}
        {\cal G}_{R} 
              & \simeq & 
                   \frac{ A \Gamma(1 + \xi)}{ M - u \cdot p } 
                   {\left[ 
                       \frac{\mu}{ i ( M - u\cdot p )} 
                   \right]}^{\xi}
       \label{off-shell} \;,
               \\
        \xi   & = & \alpha \frac{3- \lambda}{2\pi} 
       \label{param} \;,
               \\
        A     & = &
                   \exp \frac{\alpha}{\pi}
                   \left\{ 
                        1 + \frac{ 3 -\lambda }{4}
                        \left( 
                            \gamma
                            - \ln \pi 
                            - i \pi 
                        \right)
                   \right\} \;,
       \label{fact}    
\end{eqnarray}
where $\Gamma$ is the gamma function.
When $\lambda > 3$, $\xi < 0$ in Eq. (\ref{off-shell}). 
This is just as in vacuum theory\cite{Bog}, which we do not 
discuss any further. 
When $\lambda = 3$, ${\cal G}_{E}$ in Eq. (\ref{off-shell}) behaves as 
the bare propagator.

We are now in a position to discuss the decaying behavior of 
the heavy muon.
For this purpose, we should study ${\cal G}_{R}$ near on 
the muon mass shell $|M - p \cdot u| \sim 0$.
The time dependence of the heavy-muon two-point function is 
obtained through the Fourier transformation;
\begin{eqnarray}
        G_{R} & = & 
                   \left. 
                       \int \frac{d p_{0}}{2 \pi} 
                       \; {\cal G}_{R} \; \exp[ -ip_{0} t]
                   \right.
     \nonumber \\
              & \simeq & 
                       \frac{M}{E} \exp[ - iEt] 
                       \exp\biggl[ 
                              \frac{\alpha}{2\pi^2}
                              {\cal D}_{R}(\tau=M t/E , \, u)
                           \biggr] \;.
       \label{long}
\end{eqnarray}
In obtaining Eq. (\ref{long}), use has been made of 
the on-shell approximation $(M + \vec{u} 
\cdot \vec{p})/u_{0} \simeq E$.
From Eq. (\ref{long}) and Eqs. (\ref{R1})--(\ref{R3}), 
it is straightforward to obtain the time dependence 
of $G_{R}$:\\ 

\noindent a) $\tau \ll \tau_{A}$
\begin{equation}
G_{R} \simeq 
             \frac{M}{E} e^{-i E t}
             \left(
                 \frac{ \mu M t }{\sqrt{\pi} E}
             \right)^{\xi}
             \exp \left[
                      \frac{\alpha}{\pi} 
                         \left\{  
                             1 
                             + \frac{3 - \lambda}{4} 
                             ( \gamma - i \pi )               
                         \right\}
                  \right] 
  \label{longtime1}
\end{equation}
\noindent b) $\tau_{A} \ll \tau \ll \tau_{B}$
\begin{eqnarray}
        G_{R} 
            & \simeq & 
                      \frac{M}{E} e^{-i E t}
                      \left(
                          \frac{MTt}{E}
                      \right)^{\alpha \frac{MT}{E |\vec{u}|} t}
                      \exp \Biggl[
                               \alpha \frac{M T t}{E} 
                                  \left\{
                                     \frac{1}{|\vec{u}|} 
                                     \left( 
                                        - \frac{i\pi}{2} + \gamma -1 
                                     \right) 
                                     - \frac{|\vec{u}|}{2} ( \lambda -1)  
                                  \right\}
        \nonumber \\
           &        &
                   + \frac{\alpha}{\pi}
                   \left\{
                       \frac{3 - \lambda}{2}
                       \left(
                           \frac{\gamma}{2} 
                           - \ln \frac{2 \sqrt{\pi} T}{\mu}
                       \right)
                       - \frac{u_{0}}{2 |\vec{u}|} 
                       {\ln \frac{u_{0}-|\vec{u}|}{u_{0}+|\vec{u}|}}   
                       -i\pi \frac{u_{0}}{2|\vec{u}|}
                   \right\}
                          \Biggr]
  \label{longtime2}
\end{eqnarray}
\noindent c) $\tau_{B} \ll \tau$
\begin{eqnarray}
        G_{R} 
              & \simeq &
                   \frac{M}{E} e^{-i E t + F(t)} \;,
  \label{longtime4}  \\
        F(t) 
              & \equiv & 
                     - \alpha \frac{M T t}{E} |\vec{u}| 
                     \left\{ \ln \frac{M T t}{E}
                             - \frac{i\pi}{2} + \gamma -1
                             - \frac{{u_{0}}^2}{|\vec{u}|^2}
                             \left( \ln{ \frac{T}{|\vec{u}|m_{sc}}} 
                                    - \frac{i\pi}{2} 
                             \right)
                             + \frac{\lambda -1}{2} 
                     \right\} 
        \nonumber \\
              &        &
                   - \frac{\alpha}{\pi} 
                   \left\{
                       \pi
                       T \frac{{u_{0}}^2}{|\vec{u}|^2} \frac{1}{m_{sc}}
                       - \frac{3 - \lambda}{2}
                       \left(
                           \frac{\gamma}{2} 
                           - \ln \frac{2 \sqrt{\pi} T}{\mu}
                       \right)
                       + \frac{u_{0}}{2|\vec{u}|} 
                       {\ln \frac{u_{0}-|\vec{u}|}{u_{0}+|\vec{u}|}}   
                       + i \pi \frac{u_{0}}{2 |\vec{u}|} 
                   \right\} \;.                   
  \label{longtime3}
\end{eqnarray}
\narrowtext

We discuss two cases separately. 
It is to be noted that $|\vec{u}| \equiv \frac{1}{M} |\vec{p}| = 
\frac{|\vec{v}|}{\sqrt{1-{|\vec{v}|}^2}}$ with $\vec{v}$ 
being the external muon velocity:
\begin{itemize}
\item $0<v<1$\\
In the region a) above, the behavior of $G_{R}$ is gauge-parameter 
dependent, and we cannot draw any physically sensible conclusion.
In the region b), the factor ${\displaystyle 
\left( \frac{MTt}{E} \right)^{\alpha \frac{MT}{E |\vec{u}|} t}}$ 
in $G_{R}$, Eq. (\ref{longtime2}), dominates over others.
Then, the leading part of $G_{R}$ is gauge-parameter independent and 
increases as $t$ increases.
In the region c) the dominant part of $G_{R}$ is gauge-parameter 
independent, which first increases and then decreases.
The turning point in $t$ is approximately given by 
$\frac{E}{M T}(\frac{T}{m_{sc}|\vec{u}|})^{1/v^2}$. 

\item $v \simeq 1$\\
In the region a) and b) the situation is the same as 
the first sentence in the above case.
In the region c) $G_{R}$ decreases monotonously. 
\end{itemize}

Let us discuss the large $t$ region c) more elaborately.
The leading contribution comes from logarithmic 
parts of $F(t)$ in Eq. (\ref{longtime3}) (cf. Eq. (\ref{R3})). 
It is to be noted that the gauge-parameter dependent part 
in Eq. (\ref{longtime3}) may be neglected and Eq. (\ref{longtime3}) 
is gauge independent:
\begin{equation}
         F(t) 
              \simeq 
                         - \alpha \frac{MTt}{E} |\vec{u}|
                         \left( 
                             \ln \frac{M T t}{E}  
                             -  \frac{{u_{0}}^2}{|\vec{u}|^2} 
                             \ln \frac{T}{m_{sc} |\vec{u}|}
                         \right)
       \label{leading} \;.
\end{equation}

A striking feature is that Eq. (\ref{leading}) is divergence free or 
finite.
The damping rate $\gamma_{d}$ is usually defined through the behavior of 
$G_{R}$ at large $t$;
$G_{R} \sim e^{ -i E t - \gamma_{d} t} 
\;(t \longrightarrow \infty)$.
Ware it not for the factor $\ln t$ in Eq. (\ref{leading}), 
the damping rate would be identified from Eq. (\ref{longtime4}) 
with Eq. (\ref{leading}).
Eq. (\ref{leading}) indicates that, at large $t$, the heavy muon 
in a QED plasma decays much faster than the {\em constant} 
damping rate.
In any case, the quantity, $F(t)$, in Eq. (\ref{leading}) governs 
the damping behavior of the heavy muon.
The behavior of $F(t)$ in Eq. (\ref{leading}) is as follows:
\begin{itemize}
\item $0<v<1$\\
$F(t)$ decreases after  $t=t_{d} \,( \equiv \frac{E}{M T} 
( \frac{T \sqrt{1-v^2}}{m_{sc} v} )^{1/v^2})$.
This means that after $t_{d}$, the heavy muon begins to decay.

\item $v \simeq 1$\\
In this case,
\begin{equation}
F(t)  \simeq  - \alpha \frac{MTt}{E \sqrt{1 -v^2}}
            \ln \frac{M m_{sc} t}{E \sqrt{1-v^2}} 
            \;.
       \label{leading3}
\end{equation}
This is always negative and the heavy muon decays.
\end{itemize}


\section{Summary and Discussion}
In this paper we have derived and discussed the two point function 
of a heavy muon passing through a QED plasma within 
the Bloch Nordsieck approximation. 
The assumption of heavy muon means that the muon line is hard.
The electron sector only comes in as thermal radiative corrections 
to photon lines.
Among those, we only take into account the HTL corrections to 
soft photon lines.
HTL corrections to vertices are not necessary.

It is well known that, to leading order of HTL resummation scheme, 
the damping rate of a quark (muon) diverges due to the 
Coulomb-type infrared singularity even to the lowest non-trivial 
order of resummed perturbation theory.
It is because only the electric sector of soft gluon (photon) line 
is screened by the thermal mass.  
As far as QCD is concerned, the infrared singularity is screened 
by the magnetic mass which is expected to arise through 
still higher order resummations. 
But in QED, no magnetic mass is induced.

In this paper, within the Bloch Nordsieck approximation, 
it is shown that the \lq\lq damping rate\rq\rq$\,$of a heavy muon 
in a hot QED plasma is finite.
More precisely, the heavy muon decays as 
$ t^{-\alpha \frac{MT}{E} |\vec{u}| t}$ at large $t$.

Finally it is worth mentioning the case where the HTL corrections 
to the soft photon line are ignored. 
(Explicit computation is given in
Appendix B).
For $|M - p \cdot u| \gg M$, the gauge parameter dependent parts 
of ${\cal G}_{R}$ (cf. Eq. (\ref{g-r})) is the same as 
Eq. (\ref{param}).
Also for $\lambda = 3$, ${\cal G}_{R}$ reduces essentially to 
the bare propagator.
In the nearly on the mass shell case the time dependence of $G_{R}$ 
is quite different from that of 
Eq. (\ref{longtime4}) with Eq. (\ref{leading}):
\begin{equation}
         G_{R} 
              \simeq 
                     \frac{M}{E} e^{-i E t}
                     \exp 
                     \left(
                         \frac{M T }{E} t 
                     \right)^{\frac{\alpha MTt }{E|\vec{u}|}}    
       \label{leading2} \;,
\end{equation}
the anti-damping behavior!
From this fact, we learn that the HTL resummation for soft photon 
lines plays an crucial role for obtaining a physically sensible 
damping behavior.


\section*{Acknowledgement}
I gratefully thank Prof. A.~Ni\'{e}gawa for his kind help 
and thank many volunteers who develop public domain softwares. 

\appendix


\section{calculation of $D(\tau, \, u_{E})$}

According to the HTL resummation scheme, when the external photon 
lines are soft $l_{4},|\vec{l}| \sim O(eT)$, the HTL corrections 
are necessary.
Since $l_{4}$ is $2 \pi n T \; (n = 0, \pm1, \pm2, \cdots)$, 
$l_{4}$ with $n \neq 0$ is no longer in the soft region.
Then we should take care of the $n=0$ mode only.

The explicit expression for $\Pi_{T,L}$ in Eq. (\ref{pol}), within 
the HTL approximation, are given in \cite{Pis1,wel1}.
From these expression, we have
\begin{equation}
        \Pi_{T}(n=0)=0, \;\;
        \Pi_{L}(n=0)= {m_{sc}}^2 = \frac{e^2 T^2}{3} \;\;.
  \label{pi-t2}
\end{equation}
We now decompose ${\cal D}(\tau, u_{E})$ in Eq. (\ref{G7}) 
into two parts:
\begin{eqnarray}
         {\cal D}(\tau, u_{E})
             & \equiv &
                      - T (2\pi \mu)^{\epsilon} 
                      \sum_{l_{4}=even} \int d^{\, D} l 
                      \;{\cal D}(\tau, u_{E};l_{4}, \vec{l})
      \nonumber \\
             & \equiv &
                      - T (2\pi \mu)^{\epsilon} \int d^{\, D} l
                      \;{\cal D}(\tau, u_{E};l_{4}, \vec{l})
      \nonumber \\
             &        &
                      - T (2\pi \mu)^{\epsilon}
                      \sum_{l_{4}=even}
                      \raisebox{.2em}{\mbox{\hspace*{-1.8ex}}$'$} 
                      \;\; \int d^{\, D} l 
                      \;{\cal D}(\tau, u_{E};l_{4}, \vec{l})
               \;,
  \label{calc1}
\end{eqnarray}
where the prime on the summation symbol in Eq. (\ref{calc1}) 
means to take summation over $l_{4}$ with non-zero modes.
We further decompose Eq. (\ref{calc1}) as
\begin{eqnarray}
         - T (2\pi \mu)^{\epsilon} \int d^{\, D} l 
         \; {\cal D}(\tau, u_{E};l_{4}, \vec{l})
             & \equiv &
                      \sum_{m=1}^{3}
                      {\cal D}_{m}
               \;,
  \label{calc2} \\ 
         - T (2\pi \mu)^{\epsilon} 
         \sum_{l_{4}=even}
         \raisebox{.2em}{\mbox{\hspace*{-1.8ex}}$'$} 
         \;\; \int d^{\, D} l
         \; {\cal D}(\tau, u_{E};l_{4}, \vec{l})
             & \equiv &
                      \sum_{m=1}^{2}
                      {{\cal D}'}_{m}
  \label{calc3} \;,
\end{eqnarray}
where
\begin{eqnarray}
        {\cal D}_{1}
           & \equiv & 
                   \left. 
                       - T (2 \pi  \mu)^{\epsilon} \int d^{\, D}l \;
                       \frac{1 - \cos(\vec{l} \cdot \vec{u} \tau )
                            }{
                               ( \vec{l} \cdot \vec{u} )^2 } \;
                       \frac{(u_{4})^2}{ {\vec{l}}^2 +  {m_{sc}}^2 }  
                   \right. 
       \label{part1} \;, \\
        {\cal D}_{2}
           & \equiv & 
                   \left. 
                       - T (2 \pi  \mu)^{\epsilon} \int d^{\, D}l \;
                       \frac{1 - \cos(\vec{l} \cdot \vec{u} \tau )
                            }{
                              ( \vec{l} \cdot \vec{u} )^2 } \;
                       \frac{ {\vec{u}}^2 }{ {\vec{l}}^2 }    
                   \right. 
       \label{part2}  \;, \\
        {\cal D}_{3}
           & \equiv & 
                   \left. 
                       - T (\lambda -1)
                       (2 \pi  \mu)^{\epsilon} \int d^{\, D}l \;
                       \frac{1 - \cos(\vec{l} \cdot \vec{u} \tau )
                            }{
                              ( \vec{l} \cdot \vec{u} )^2 } 
                       {\left( 
                            \frac{ \vec{l} \cdot \vec{u} 
                                }{ {\vec{l}}^2 } 
                       \right)}^2 
                   \right. 
       \label{part3} \;, \\
        {{\cal D}'}_{1}
           & \equiv & 
                  \left. 
                      -T (2 \pi \mu)^{\epsilon}
                      \sum_{l_{4}=even}
                      \raisebox{.2em}{\mbox{\hspace*{-1.8ex}}$'$} 
                      \;\; \int d^{\, D}l \;
                      \frac{1 - \cos(l \cdot u_{E} \tau )
                           }{
                             ( l \cdot u_{E} )^2 } \;
                      \frac{ {u_{E}}^2 }{ l^2 }  
                  \right. \;, 
       \label{partl1} \\
        {{\cal D}'}_{2}
           & \equiv & 
                 \left. 
                     -T(\lambda -1) (2 \pi \mu)^{\epsilon}
                     \sum_{l_{4}=even}
                     \raisebox{.2em}{\mbox{\hspace*{-1.8ex}}$'$} 
                     \;\; \int d^{\, D}l \;
                     \frac{1 - \cos(l \cdot u_{E} \tau )
                          }{
                            ( l \cdot u_{E} )^2 } 
                 \right.
      \nonumber \\
             &        &
                     \times
                     \left( 
                         \frac{ l \cdot u_{E} }{ l^2 }      
                     \right)^2 
                     \;. 
       \label{partl3}
\end{eqnarray} 
It is to be noted that the integrand of ${\cal D}_{m} , \; (m=1,2,3)$ 
and ${{\cal D}'}_{m} , \; (m=1,2)$ are regular at $\vec{l}= \vec{0}$.

We first evaluate ${\cal D}_{m} , \; (m=1,2,3)$. 
Carried out the integration over the polar angle 
$ \cos^{-1}(\frac{\vec{l} \cdot \vec{u}}{|\vec{l}||\vec{u}|})$,
Eq. (\ref{part1}) becomes
\widetext
\begin{equation}
{\cal D}_{1} 
      =
        - T (2 \pi  \mu)^{\epsilon} 
        \frac{{u_{4}}^2}{|\vec{u}|^2} 
        2 \pi^{1- \epsilon / 2} \Gamma (\epsilon /2)
        \int_{0}^{\infty} d l \;       
        (l^2 + m_{sc}^2 )^{-\epsilon}
        \left\{ 
            \frac{1 - e^{i |\vec{u}| \tau l}}{l^2} + c.c.
        \right\}    
  \label{polar1} \;.
\end{equation}
In order to perform the integration over $l$, 
it is convenient to analytically continue the integrand into a complex 
$l$ plane.
By deforming the integration contour (as shown in Fig. \ref{fig1}), 
we obtain, 
\begin{eqnarray}
        {\cal D}_{1}
              & = & 
                   - T (2 \pi  \mu)^{\epsilon} 
                   \frac{{u_{4}}^2}{|\vec{u}|^2} 
                   2 \pi^{1- \epsilon / 2} \Gamma (\epsilon /2)
                   \Biggl[ 
                        \frac{\pi \tau |\vec{u}|}{2} 
                        m_{sc}^{-\epsilon}
                        - \frac{i}{2} \int_{m_{sc}}^{\infty} d l
                        \frac{(l^2 - {m_{sc}}^2 )^{-\epsilon /2}}{l^2}
                        e^{-i \frac{\pi}{2} \epsilon}
                        ( 1 - e^{- |\vec{u}| \tau l} ) + c.c.
     \nonumber \\              
              &   &   
                        - i \lim_{\rho \longrightarrow 0} 
                        \int_{\rho}^{m_{sc}} d l
                        \frac{(l^2 - {m_{sc}}^2 )^{-\epsilon /2}}{l^2}
                        ( 1 - e^{- |\vec{u}| \tau l} ) + c.c. 
                   \Biggr]
     \nonumber \\              
              & = &   
                   - 2 \pi^2 T \tau \frac{{u_{4}}^2}{|\vec{u}|}
                   \left(
                       \frac{1}{\epsilon}
                       + \ln \frac{2\pi \mu }{\sqrt{\pi} m_{sc}} 
                       - \frac{\gamma}{2}    
                       - \frac{1 - e^{- |\vec{u}| \tau m_{sc}}
                              }{
                             |\vec{u}| \tau m_{sc}
                               }
                       + E_{i}(- |\vec{u}| \tau m_{sc})
                   \right)   
     \nonumber \\              
              & = &   
                   -2 \pi^2 T \tau 
                   \frac{{u_{4}}^2}{|\vec{u}|}
                   \left( 
                       \frac{1}{\epsilon} 
                       + \ln \frac{ 2 \pi \mu}{\sqrt{\pi} m_{sc}}
                       - \frac{\gamma}{2}  
                   \right) 
        \nonumber \\
              &   &  
                   + 2 \pi^2 T \frac{{u_{4}}^2}{{\vec{u}}^2}
                   \left\{
                   \begin{array}{ll}
                           |\vec{u}| \tau 
                           (1- \gamma - \ln |\vec{u}| \tau m_{sc})
                           + O(|\vec{u}|^2 \tau^2 m_{sc})
                           & (|\vec{u}| \tau m_{sc} \ll 1) \\
                           &                                   \\
                           {\displaystyle \frac{1}{m_{sc}}} 
                           + \left.
                               O(\frac{e^{- |\vec{u}| \tau m_{sc} } 
                                      }{m_{sc}})
                             \right.
                           & (|\vec{u}| \tau m_{sc} \gg 1) 
                   \end{array}
                   \right. \;.
       \label{part4} 
\end{eqnarray}
\narrowtext
Through similar procedures, ${\cal D}_{2}$ and ${\cal D}_{3}$ 
may be evaluated: 
\begin{eqnarray}
        {\cal D}_{2}
              & = & 
                   -2 \pi^2 T |\vec{u}| \tau 
                   \left( 
                       \frac{1}{\epsilon}
                       + \ln \frac{2\pi\mu|\vec{u}| \tau}{\sqrt{\pi}}
                       +\frac{\gamma}{2} -1 
                   \right) 
       \label{part5} \\
        {\cal D}_{3}
              & = &
                   - (\lambda-1) \pi^2 T |\vec{u}| \tau 
       \label{part6} \;.
\end{eqnarray}
The terms proportional to $1/\epsilon$ , which correspond to UV 
divergent piece at $\epsilon=0$ or $D=3$, is to be eliminated later 
by renormalization.

Let us evaluate ${{\cal D}'}_{m} \; (m=1,2)$ in Eq. (\ref{calc3}).
To evaluate both Eq. (\ref{partl1}) and (\ref{partl3}), 
it is convenient to introduce 
\begin{equation}
         F(l_{4})
                \equiv
                      -T (2 \pi \mu)^{\epsilon}
                      \int d^{\, D}l \;
                      \frac{1 - \cos(l \cdot u_{E} \tau )
                           }{
                             ( l \cdot u_{E} )^2 }\; 
                      \frac{ 1 }{ l^2 + \sigma }  
                   \;.
  \label{conv1}
\end{equation}
The integral in Eq. (\ref{partl1}) is obtained from $F(l_{4})$ 
with $\sigma = 0$, while the integral in Eq. (\ref{partl3}) 
is obtained from $\partial F(l_{4}) / \partial \sigma$ with 
with $\sigma = 0$.
Carrying out the integrations over the polar angle and using 
the technique of deforming the integration contour 
in the $l$-plane, we obtain 
\begin{eqnarray}
         F(l_{4})
            & = &
                 -T (2 \pi \mu)^{\epsilon}
                 \pi^{1 - \epsilon /2} \Gamma (\epsilon /2)
                 \Biggl[
                     \frac{\pi \tau}{|\vec{u}|}
                     \left\{
                          l_{4}^{2} 
                          \left(
                              1 + \frac{{u_{4}}^2}{|\vec{u}|^2}
                          \right)
                          + \sigma
                     \right\}^{-\epsilon /2}
      \nonumber \\              
            &   & 
                 + \sin \frac{\pi \epsilon}{2} 
                 \int_{\sqrt{l_{4}^2 + \sigma}}^{\infty} d l \;  
                 \left\{
                     l^2 - (l_{4}^2 + \sigma)
                 \right\}^{-\epsilon /2}
      \nonumber \\              
            &   & 
                 \times
                 \left\{
                     \frac{1}{(u_{4} l_{4} + i |\vec{u}| l)^2 }
                     + c.c 
                     - \frac{e^{i u_{4} l_{4} \tau - |\vec{u}| l \tau}
                          }{
                          (u_{4} l_{4} + i |\vec{u}| l)^2
                           }
                     + c.c 
                 \right\}
                 \Biggr] \;.
      \nonumber \\              
            &   & 
  \label{conv2}
\end{eqnarray}
As to the integration of the third term in last brackets 
in Eq. (\ref{conv2}), we first carry out the integration by parts 
and the integration by deforming the contour $C_{1}$ to $C_{2} 
\bigoplus C_{3}$ in Fig. 2.
Then we shift the integration valiable to get 
\begin{eqnarray}
         F(l_{4})
            & = &
                 -T (2 \pi \mu)^{\epsilon}
                 \pi^{1 - \epsilon /2} \Gamma (\epsilon /2)
                 \Biggl[
                     \frac{\pi \tau}{|\vec{u}|}
                     \left\{
                          l_{4}^{2} 
                          \left(
                              1 + \frac{{u_{4}}^2}{|\vec{u}|^2}
                          \right)
                          + \sigma
                     \right\}^{-\epsilon /2}
       \nonumber \\              
             &   & 
                 + \sin \frac{\pi \epsilon}{2} 
                 (l_{4}^2 + \sigma)^{- \frac{1+ \epsilon}{2}}
                 \int_{0}^{\infty} d l \;
                 \displaystyle{
                       \frac{ (l^2 -1)^{-\epsilon /2}}{
                             \left(
                                 i l |\vec{u}| 
                                 + \frac{l u_{4}}{\sqrt{l_{4}^2
                                        + \sigma }}
                             \right)^2
                            }
                             }                           
                 + c.c
      \nonumber \\              
            &   &     
                 + \frac{\tau}{|\vec{u}|}
                 \int_{\sqrt{l_{4}^2 + \sigma}}^{\infty} d l \;
                 \frac{e^{-l |\vec{u}| \tau}}{l}
                 + \frac{1}{|\vec{u}|}
                 \frac{e^{i l_{4} u_{4} \tau 
                          - |\vec{u}| \tau 
                          \sqrt{l_{4}^2 + \sigma}}
                     }{
                       i l_{4} u_{4} - |\vec{u}| 
                       \sqrt{l_{4}^2 + \sigma}
                     }
                 + c.c.
      \nonumber \\              
            &   &     
                 + \frac{i \tau}{|\vec{u}|}
                 \int_{0}^{\frac{l_{4} u_{4}}{|\vec{u}|}} d l \;
                 \frac{e^{- |\vec{u}| \tau 
                          (\sqrt{l_{4}^2 + \sigma} - i l)}
                     }{
                       \sqrt{l_{4}^2 + \sigma}
                       - i l
                     }
                 + c.c. 
                 \Biggr] \;.
  \label{conv3}
\end{eqnarray}
\widetext
Using Eq. (\ref{conv3}) with $\sigma = 0$ and carrying out 
the summation over $l_{4}$, we obtain 
\begin{eqnarray}
        {{\cal D}'}_{1}
            & = &  
                    \sum_{l_{4}=even}
                    \raisebox{.2em}{\mbox{\hspace*{-1.8ex}}$'$} 
                    {u_{E}}^2 F(l_{4})\Biggr|_{\sigma =0}                    
      \nonumber \\              
            & = &                       
                    2\pi^2 T \tau\frac{{u_{E}}^2}{|\vec{u}|} 
                    \left(
                       \frac{1}{\epsilon} -\frac{\gamma}{2}
                       -\ln \sqrt{\pi} 
                    \right)
                    + \frac{2\pi}{\epsilon} +\pi \gamma
                    - 2\pi \ln \frac{4\pi^{3/2} T}{2\pi \mu}
                    - \frac{2\pi u_{4}}{|\vec{u}|} 
                    \tan^{-1} \frac{u_{4}}{|\vec{u}|} 
      \nonumber \\              
              &   & 
                       - 2 \pi^2 T \tau \frac{{u_{E}}^2}{|\vec{u}|}
                       \left(
                           \ln \frac{T}{2\pi \mu}
                           + \ln \frac{u_{E}}{|\vec{u}|} 
                       \right)
                       + 4 \pi^2 T \tau \frac{{u_{E}}^2}{|\vec{u}|}
                       \int_{1}^{\infty} d x \; \frac{1}{x}
                       \frac{1}{
                              e^{2\pi T |\vec{u}| \tau x} -1 
                               }
      \nonumber \\              
              &   & 
                       + \frac{ \pi {u_{E}}^2 
                              }{  |\vec{u}| 
                             (|\vec{u}|  -i u_{4}) 
                              } 
                       \ln (1 - 
                          e^{-2 \pi T \tau (|\vec{u}| -i u_{4})} 
                           ) + c.c. 
      \nonumber \\              
              &   & 
                       + 2 \pi^2 T \tau 
                       \frac{u_{4} {u_{E}}^2}{|\vec{u}|} 
                       \left(
                           i \int_{0}^{1} d x \;
                           \frac{1}{
                             |\vec{u}| - i u_{4} x  
                                   }\;
                           \frac{1}{
                              e^{2 \pi T \tau (|\vec{u}| 
                                  - i u_{4} x )
                                }
                               - 1 }
                           + c.c.
                       \right) \;.
\end{eqnarray}

${{\cal D}'}_{2}$ in Eq. (\ref{partl3}) may be evaluated as:
\begin{equation}
        {{\cal D}'}_{2}
               = 
                 - ( \lambda - 1) 
                 \sum_{l_{4}=even}
                 \raisebox{.2em}{\mbox{\hspace*{-1.8ex}}$'$} 
                 \;\;   
                 \Biggl[  
                      T ( 2\pi \mu)^{\epsilon}  
                      \int d^{\, D} l
                      \frac{1}{l^4}
                      - \frac{\partial^{\, 3}
                             }{
                                \partial \sigma \partial \tau^2
                              }
                      F(l_{4}) \Biggr|_{\sigma=0}
                 \;\;
                 \Biggr]  
                 \;, 
  \label{conv4} 
\end{equation}
and we obtain,
\begin{eqnarray} 
        {{\cal D}'}_{2}
            & = &  
                  - (\lambda -1) \pi 
                   \left[
                       + \frac{1}{ \epsilon} 
                       + \ln \frac{2 \pi\mu}{\sqrt{\pi}}
                       - \gamma
                       + (\gamma - \ln 4 \pi T)
                       + 2 \ln (1 -e^{-2 \pi T |\vec{u}| \tau })
                       + 2 |\vec{u}| \tau \pi T
                       \frac{1}{e^{ 2 \pi T |\vec{u}| \tau} -1}
                  \right.
      \nonumber \\              
              &   &
                       - \frac{1}{2} \ln ( 1 
                          - e^{- 2 \pi T \tau 
                          (|\vec{u}| - i u_{4}) 
                              }           
                                         ) 
                       - c.c.
      \nonumber \\              
              &   &
                       - \pi T \tau
                          (|\vec{u}| - i u_{4}) 
                          \frac{1}{
                            e^{2 \pi T \tau 
                              ( |\vec{u}| - i u_{4} ) -1 }
                                  }
                       - c.c.
      \nonumber \\              
              &   &
                       - 3 i u_{4} \pi T \tau
                       \int_{0}^{1} dx
                       \frac{1}{
                          e^{2 \pi T \tau 
                          (|\vec{u}| -i u_{4} ) 
                            } -1 
                                }
                       + c.c.
      \nonumber \\              
              &   &
                       + i 2 \tau^2 \pi^2 T^2 u_{4}                            
                       \int_{0}^{1} dx \;
                       (|\vec{u}| -i u_{4} x) 
                       \frac{
                        e^{ -2 \pi T \tau
                        (|\vec{u}| -i u_{4} x)  } 
                           }{
                        ( e^{ -2 \pi T \tau
                        (|\vec{u}| -i u_{4} x) } -1 
                        )^2
                            }
                       + c.c. 
                   \Biggr] \;. 
\end{eqnarray}

Now let us study the behaviors of ${{\cal D}'}_{m} \; (m=1,2)$ in the 
following two limits ($\tau_{A}$ is as in the text, 
$\tau_{A}= \frac{1}{|\vec{u}| T}$):\\

\noindent a) $\tau  \ll \tau_{A}$
\begin{eqnarray}
        {{\cal D}'}_{1}
            & \simeq &  
                 2 \pi 
                 \left( 
                     \frac{1}{\epsilon} + \frac{\gamma}{2} +1 
                     + \ln \frac{2 \pi \mu u_{E} \tau}{2 \sqrt{\pi}}
                 \right) 
                 + 2\pi^2 \tau T \frac{{u_{E}}^2}{|\vec{u}|} 
                 \left( 
                     \frac{1}{\epsilon} + \frac{\gamma}{2} -1 
                     + \ln \frac{2 \pi \mu |\vec{u}| \tau
                                }{ \sqrt{\pi}}
                     \right) \;,
        \nonumber \\
       \label{partl4} \\
        {{\cal D}'}_{2}
            & \simeq & 
                 -(\lambda-1) \pi 
                 \left( 
                     \frac{1}{\epsilon} + \frac{\gamma}{2} +1 
                     + \ln \frac{2 \pi \mu u_{E} \tau
                                }{ 2  \sqrt{\pi}} 
                 \right) 
                 + (\lambda -1) \pi^2 T |\vec{u}| \tau 
       \label{partl5} \;,
\end{eqnarray}
and,\\
\\
b) $\tau  \gg \tau_{A}$
\begin{eqnarray}
        {{\cal D}'}_{1}
            & \simeq &   
                  2 \pi 
                  \left( 
                      \frac{1}{\epsilon} +\frac{\gamma}{2}
                      - \ln \frac{4\pi^{3/2}T}{2\pi\mu}
                      - \frac{u_{4}}{|\vec{u}|}
                      \tan^{-1} \frac{u_{4}}{|\vec{u}|} 
                  \right)
                  + 2 \pi^2 T \tau \frac{{u_{E}}^2}{|\vec{u}|}
                  \biggl( 
                      \frac{1}{\epsilon} - \frac{\gamma}{2}
                      + \ln \frac{2 \pi \mu |\vec{u}|}{
                               u_{E} \sqrt{\pi} T}
                  \biggr) \;,
       \label{partl6} \\
       {{\cal D}'}_{2}
            & \simeq & 
                 - (\lambda -1) \pi 
                 \left(
                     \frac{1}{\epsilon}+ \frac{\gamma}{2}
                     - \ln \frac{4 \pi^{3/2} T}{2\pi \mu} 
                 \right) \;.
       \label{partl7}
\end{eqnarray}

Substituting all the results obtained above into Eq. (\ref{calc1}), 
we deduce the following behaviors of ${\cal D}(\tau, u_{E})$ 
in various limits ($\tau_{B}$ is as in the text, 
$\tau_{B}= \frac{1}{m_{sc}|\vec{u}|}$):\\

\noindent a)  $ \tau  \ll \tau_{A}$
\begin{eqnarray}
        {\cal D}(\tau, u_{E})
            & \simeq & 
                 2 \pi 
                 \left( 
                     \frac{1}{\epsilon} + \frac{\gamma}{2} + 1 
                     + \ln \frac{ 2\pi \mu \tau u_{E}
                                }{ 2 \pi \sqrt{\pi}}
                 \right)
                 - \pi (\lambda -1) 
                 \left( 
                     \frac{1}{\epsilon} + \frac{\gamma}{2} 
                     + \ln \frac{ 2\pi \mu \tau u_{E} 
                                }{ 2\pi \sqrt{\pi}} 
                 \right) 
       \label{bare1} \;,
\end{eqnarray}
\\
b)  $ \tau_{A} \ll \tau  \ll \tau_{B} $
\begin{eqnarray}
        {\cal D}(\tau, u_{E})
            & \simeq & 
                 \left.              
                     - 2 \pi^2 T \tau \frac{{u_{E}}^2}{|\vec{u}|}
                     ({\ln \tau T u_{E}} + \gamma -1) 
                 \right.
                 + 2 \pi 
                 \left( 
                     \frac{1}{\epsilon} + \frac{\gamma}{2} 
                     - \ln \frac{ 4\pi^{3/2} T}{2\pi \mu} 
                     - \frac{u_{4}}{|\vec{u}|} 
                     \tan^{-1}\frac{u_{4}}{|\vec{u}|} 
                \right)
       \nonumber \\
            &   &    
                 - (\lambda -1) \pi^2 T |\vec{u}| \tau 
                 - (\lambda -1) \pi 
                 \left(
                     \frac{1}{\epsilon} + \frac{\gamma}{2}   
                     - \ln \frac{ 4\pi^{3/2} T}{2\pi \mu}
                     \right)
       \label{bare2} \;,
\end{eqnarray}
\\
c) $\tau_{B} \ll \tau  $
\begin{eqnarray}
        {\cal D}(\tau, u_{E})
            & \simeq & 
                 \left.     
                     - 2 \pi^2 T \tau \frac{{u_{E}}^2}{|\vec{u}|}
                     ({\ln \tau T u_{E}} + \gamma -1) 
                 \right.
                 + 2 \pi 
                 \left( 
                     \frac{1}{\epsilon} + \frac{\gamma}{2} 
                     - \ln \frac{ 4\pi^{3/2} T}{2\pi \mu} 
                     - \frac{u_{4}}{|\vec{u}|} 
                     \tan^{-1} \frac{u_{4}}{|\vec{u}|} 
                 \right)
        \nonumber \\
            &   &    
                 - (\lambda -1) \pi^2 T |\vec{u}| \tau 
                 - (\lambda -1) \pi 
                 \left(
                     \frac{1}{\epsilon} + \frac{\gamma}{2} 
                     - \ln \frac{ 4\pi^{3/2} T}{2\pi \mu}  
                 \right)
        \nonumber \\
            &   &   
                 \left.     
                     - 2 \pi^2 T |\vec{u}| \tau 
                     \frac{{u_{4}}^2}{|\vec{u}|^2}
                     ( -\gamma +1 -\ln{|\vec{u}| \tau m_{sc}} ) 
                     + 2 \pi^2 T \frac{{u_{4}}^2}{|\vec{u}|^2}
                     \frac{1}{m_{sc}} 
                 \right. \;.
       \label{bare3}
\end{eqnarray}
\narrowtext
The remaining tasks are to eliminate UV divergence through 
renormalization with MS scheme and to perform 
the $\tau$ integration (see text).


\section{no HTL correction to the photon propagator}

We set $\Pi_{T,L} =0$ in Eq. (\ref{G7}).
In this case, Eq. (\ref{G7}) reduces to 
\begin{eqnarray}
        {\cal D}(\tau,u_{E})\Biggr|_{m_{sc}=0} 
            & \equiv &  
                      {\cal D}_{1+2}\Biggr|_{m_{sc}=0} 
                      + {\cal D}_{3}\Biggr|_{m_{sc}=0} 
                      + \sum_{m=1}^{2} {{\cal D}'}_{m}\Biggr|_{m_{sc}=0} 
       \label{no-HTL} \\
            &   =    &        
                 \left. 
                     - T (2 \pi \mu)^{\epsilon} \int d^{\, D}l
                     \frac{1 - \cos(\vec{l} \cdot \vec{u} \tau )
                          }{
                           ( \vec{l} \cdot \vec{u} )^2 } 
                 \right.
      \nonumber \\
            &        &        
                 \times
                 \left\{ 
                     \frac{{u_{E}}^2}{ \vec{l}^2  } 
                     + (\lambda -1) 
                     \left( 
                         \frac{ \vec{l} \cdot \vec{u} }{ \vec{l}^2 }      
                     \right)^2 
                 \right\} \;.
       \label{d-m0}
\end{eqnarray}
Here
\begin{eqnarray}
        {\cal D}_{1+2}\Biggr|_{m_{sc}=0} 
            & \equiv & 
                 \left. 
                     - T (2 \pi \mu)^{\epsilon} \int d^{\, D}l \;
                     \frac{1 - \cos(\vec{l} \cdot \vec{u} \tau )
                          }{
                           ( \vec{l} \cdot \vec{u} )^2 } 
                     \frac{{u_{E}}^2}{ \vec{l}^2 }  
                 \right. \;,
       \label{d-m01} \\
        {\cal D}_{3}\Biggr|_{m_{sc}=0} 
            & \equiv & 
                 \left. 
                     - ( \lambda -1) T (2 \pi \mu )^{\epsilon} 
                     \int d^{\, D}l \;
                     \frac{1 - \cos(\vec{l} \cdot \vec{u} \tau )
                          }{
                            ( \vec{l} \cdot \vec{u} )^2 } 
                     {\left( 
                          \frac{ \vec{l} \cdot \vec{u} }{ \vec{l}^2 }      
                     \right)}^2 
                 \right. \;.
       \label{d-m02}
\end{eqnarray}
Similar procedures as in Appendix A leads to 
\begin{eqnarray}
        {\cal D}_{1+2}\Biggr|_{m_{sc}=0} 
            & \simeq & 
                 - 2 \pi^2 \tau T \frac{{u_{E}}^2}{|\vec{u}|}
                 \left( 
                     \frac{1}{\epsilon} + \frac{\gamma}{2} -1
                     + \ln \frac{2\pi \mu |\vec{u}| \tau}{\sqrt{\pi}}
                 \right)  \;,
     \label{D-12} \\
        {\cal D}_{3}\Biggr|_{m_{sc}=0} 
            & \simeq &
                 - (\lambda -1) \pi^2 T  |\vec{u}| \tau
     \label{D-3} \;.
\end{eqnarray}
Substituting the above Eqs. (\ref{D-12}) and (\ref{D-3}) and 
Eqs. (\ref{partl4})--(\ref{partl7}) to Eq. (\ref{no-HTL}), we have
\\
\widetext
a) $\tau  \ll \tau_{A}$
\begin{equation}
        {\cal D}(\tau ,u_{E}) \Biggr|_{m_{sc}=0} 
              \simeq  
                 2 \pi 
                 \left( 
                     \frac{1}{\epsilon} + \frac{\gamma}{2} +1 
                     + \ln \frac{2 \pi \mu u_{E} \tau
                                }{ 2  \sqrt{\pi}} 
                 \right) 
                 - \pi (\lambda -1) 
                 \left( 
                     \frac{1}{\epsilon} + \frac{\gamma}{2} 
                     + \ln \frac{2 \pi \mu u_{E} \tau }{ 2 \sqrt{\pi} } 
                 \right) \;.
\end{equation}

\noindent b) $\tau  \gg \tau_{A}$ \\
\begin{eqnarray}
        {\cal D}(\tau, u_{E})\Biggr|_{m_{sc}=0} 
            & \simeq & 
                 -2 \pi^2 T \tau \frac{{u_{E}}^2}{|\vec{u}|}
                 \left( 
                     \ln T u_{E} \tau +\gamma -1 
                 \right)
                 + 2\pi 
                 \left( 
                     \frac{1}{\epsilon} + \frac{\gamma}{2}
                     - \ln \frac{4 \pi^{3/2} T}{2 \pi \mu} 
                     - \frac{u_{4}}{|\vec{u}|}
                     \tan^{-1} \frac{u_{4}}{|\vec{u}|}
                 \right)   
      \nonumber \\
            &   &
                 - \pi^2 (\lambda -1) T|\vec{u}| \tau 
                 - \pi(\lambda -1) 
                 \left( 
                     \frac{1}{\epsilon} + \frac{\gamma}{2} 
                     - \ln \frac{2 \pi \mu u_{E} \tau }{ 2 \sqrt{\pi} } 
                 \right) \;.
\end{eqnarray}
\narrowtext



\begin{figure}
\caption{Integration contour in the complex $l$ plane} 
\label{fig1}
\end{figure}

\begin{figure}
\caption{The integration along the contour $C_{1}$ may be 
evaluated along the contour $C_{2} \bigoplus C_{3}$.}
\label{fig2}
\end{figure}


\begin{references}
%
\bibitem{Blo}
F.~Bloch, A.~Nordsieck, Phys. Rev. {\bf 52}, 54 (1937).
%
%
\bibitem{Yen}
D.~R.~Yennie, S.~C.~Frautschi, and H.~Suura, Ann. Phys. (N.Y.)
{\bf 13}, 379(1961); 
T.~D.~Lee and M.~Nauenberg, Phys. Rev. {\bf 133}, B1549 (1964); 
J.~M.~Jauch and F.~Rohrlich, {\em The Theory of Photons and 
Electrons} (Springer--Verlag, 1980).
%
%
\bibitem{Mut}T.~Muta, {\em 
Foundations of Quantum Chromodynamics} (World Scientific, 
Singapore, 1987).
%
%
\bibitem{nie}
A.~Ni\'{e}gawa, K.~Takashiba, Nucl. Phys. {\bf B370}, 335 (1992):
See also M.~Le.~Bellac and P.~Reynaud, in:
{\em Banff/CAP workshop on thermal field theory},
edited by F.~C.~Khnna, R.~Kobes, G.~Kunstatter, 
and H.~Umezawa, (World Scientific, Singapore, 1994) p.440.
%
%
\bibitem{wel}
H.~A.~Weldon, Phys. Rev. D {\bf 44}, 3955 (1991); {\bf 49}, 1579 (1994).
%
%
\bibitem{Pis0}
R.~D.~Pisarski, Phys. Rev. Lett. {\bf 63}, 1129 (1989). 
%
%
\bibitem{Pis1}
E.~Braaten and R.~D.~Pisarski, Nucl. Phys. {\bf B337}, 569 (1990).
%
%
\bibitem{bur1}
E.~Braaten and M.~H.~Thoma, Phys. Rev. {\bf D44}, 1298, R2625 (1991);
J.~I.~Kapusta, P.~Lichard, and D.~Seibert, {\it ibid}. {\bf D44}, 
2774 (1991); 
R.~Baier, H.~Nakkagawa, A.~Ni\'{e}gawa, and K.~Redlich. Z. Phys. 
{\bf C53}, 433 (1992); {\it ibid}. {\bf 45}, 4323 (1992).
%
%
\bibitem{bur2}
C.~P.~Burgess and A.~L.~Marini, Phys. Rev. {\bf D45}, R17 (1992); 
A.~Rebhan, {\it ibid}. {\bf D46}, 482 (1992).
R.~Baier, H.~Nakkagawa, A.~Ni\'{e}gawa, in Proceedings of 
the Workshop on High Temperature Field Theory, Winnipeg, July 1992 
Can. J. Phys. (to be published);  
R.~D.~Pisarski, Phys. Rev. {\bf D47}, 5589 (1993); 
M.~E.~Carrington, University of Winnipeg Report No. WIN-93-xx, 
1993 (unpublished); 
T.~Altherr and D.~Seibert, CERN Report No. CERN-TH.6882/93, 1993 
(unpublished). 
See, also, V.~V.~Lebedev and A.~V.~Smilga, Ann. Phys. (N.Y.) 
{\bf 202}, 229 (1990); Physica {\bf A181}, 187 (1992); 
T.~Altherr, E.~Petitgirard, and T.~del~Rio~Gaztelurrutia, Phys. Rev. 
{\bf D47}, 703 (1993); 
A.~V.~Smilga, Bern University Report No. BUTP-92/39, 1992 
(unpublished); S.~Peign\'{e}, E.~Pilon, and D.~Schiff, Orsay Report 
No. LPTHE Orsay 93-13, 1993 (unpublished); 
R.~Baier and R.~Kobes, Phys. Rev. {\bf D50}, 5944 (1994); 
A.~Ni\'{e}gawa, Phys. Rev. Lett {\bf 73}, 2023 (1994). 
%
%
\bibitem{Bog}
N.~N.~Bogoliubov, D.~V.~Shirkov, 
{\em INTRODUCTION TO THE THEORY OF QUANTIZED FIELDS} 
(Interscience Publishers Ltd., London, 1959)
%
%
\bibitem{Ste} A.~Kernemann and N.~G.~Stefanis, 
Phys. Rev. D {\bf 40}, 2103 (1989).
%
%
\bibitem{Mat}
T.~Matsubara, Prog. Theor. Phys. {\bf 14}, 351 (1955); 
See also J.~I.~Kapusta, {\em Finite-temperatur field theory} 
(Camblidge University Press, 1989); 
N.~P.~Landsman and Ch.~G.~van~Weert, Phys. 
Rep. {\bf 145}, 141 (1987).
%
%
\bibitem{Itz}See e.g. C.~Itzykson and J.~-B.~Zuber, {\em Quantum 
Field Theory} (McGraw-Hill, New York, 1980); 
J.~D.~Bjorken and S.~D.~Drell, {\em Relativistic Quantum Fields} 
(McGraw-Hill, New York, 1965). 
%
%
\bibitem{Pis2}
E.~Braaten and R.~D.~Pisarski, Nucl. Phys. {\bf 339}, 310 (1990);  
See also, J.~Frenkel and J.~C.~Taylor, Nucl. Phys. {\bf B334}, 199 
(1990).
%
%
\bibitem{wel1}V.~P.~Silin, Zh. Eksp. Teor. Fiz. {\bf 38}, 1577 
(1960) [Sov. Phys. JETP {\bf 11}, 1136 (1960)]; 
V.~V.~Klimov, Yad. Fiz. {\bf 33}, 1734 (1981) [Sov. J. Nucl. Phys. 
{\bf 33}, 934 (1981)]; Zh. Eksp. Teor. Fiz. {\bf 82}, 336 (1982) 
[Sov. Phys. JETP {\bf 55}, 199 (1982)]; 
H.~A.~Weldon, Phys. Rev. D {\bf 26}, 1394 (1982); 
R.~D.~Pisarski, Physica {\bf 158A}, 146 (1989).
%
%
\end{references}
\end{document}